\documentclass[useAMS,usenatbib]{mn2e}

\usepackage{epsfig}
\usepackage{amsmath}
\usepackage{amssymb}

\title{The stellar scintillation on large and extremely large telescopes}
\author[V.~Kornilov]{V.~Kornilov\thanks{E-mail: victor@sai.msu.ru}\\
Lomonosov Moscow State University, Sternberg Astronomical Institute, Universitetsky prosp. 13, 119234 Moscow, Russia}

\begin{document}
\date{Accepted 2012 July 4.  Received 2012 July 3; in original form 2012 June 13}

\pagerange{\pageref{firstpage}--\pageref{lastpage}}
\pubyear{2012}

\maketitle
\label{firstpage}

\begin{abstract}
The accuracy of ground-based astronomical photometry is limited by two factors: photon statistics and stellar scintillation arising when star light passes through Earth's atmosphere. This paper examines the theoretical role of the outer scale $L_0$ of the optical turbulence (OT) which suppresses the low-frequency component of scintillation. It is shown that for typical values of $L_0 \sim 25 - 50$~m, this effect becomes noticeable for a telescopes of diameter around $4$~m. On extremely large, $30 - 40$~m, telescopes with exposures longer than a few seconds, the inclusion of the outer scale in the calculation reduces the scintillation power by more than a factor of 10 relative to conventional estimates. The details of this phenomenon are discussed for various models of non-Kolmogorov turbulence. Also, a quantitative description of the influence of the telescope central obscuration on the measured scintillation noise is introduced and combined with the effect of the outer scale. Evaluation of the scintillation noise on the future TMT and E-ELT telescopes, predicts an amplitude of approximately $10~\mu\mbox{mag}$ for a 60~s exposures.
\end{abstract}

\section{Introduction}

Stellar scintillation is the random fluctuation of the radiation flux entering the aperture of the telescope caused by amplitude distortions of light wave passing through the turbulent terrestrial atmosphere \citep{Tatarsky1967,Roddier1981}. The photometric error due to scintillation, the {\it scintillation noise}, has been repeatedly studied \citep[see, e.g.,][]{Young1969,Dravins1998}, as it often determines the fundamental limit of the accuracy of ground-based photometry.

The basic dependencies required to calculate scintillation noise have been known by the wide astronomical community for quite a while now \citep{Heasley1996,Gilliland1993}. However, these relationships were obtained for ideal cases and cannot always be used for accurate prediction of the scintillation noise. This became especially noticeable when  accurate measurements of the intensity of the optical turbulence (OT) in the atmosphere at altitudes responsible for the occurrence of scintillation on large telescopes, began to be accessible \citep{Kenyon2006}.

The content of this paper is a theoretical study of two factors affecting the power of stellar scintillation on large and extremely large telescopes: the influence of the OT outer scale and the telescope central obscuration (CO).

Section \ref{sec:theory} recalls the basic description of the phenomenon and its relationship with several parameters of the OT in the atmosphere. In the following section, we assess the influence of the outer scale in the case of  different and simplified  non-Kolmogorov models. The effect of the CO is considered in Section \ref{sec:co}, first for the Kolmogorov model, and then for a general case. In the last Section, the conclusions are formulated and a prediction of the scintillation noise for future extremely large telescopes is given.

\section{Theory}
\label{sec:theory}

\subsection{Scintillation noise in photometry}

Measurements of the brightness of astronomical objects are always burdened by noises of several different nature. Depending on the origin, these noises are included in the measured signal in different ways. Noise caused by the stellar scintillation, enters by multiplicative way, i.e. for it, the signal to noise ratio is independent of brightness. Stellar scintillation is usually characterised by {\it index} $s^2$, representing the variance of the relative fluctuations of the intensity $I$ of the radiation, passing through the receiving aperture:
\begin{equation}
s^2 = \langle (I-\langle I \rangle)^2 \rangle/\langle I \rangle^2.
\label{eq:def_index}
\end{equation}
Although in this definition, the averaging over all possible states (over ensemble) is performed, in practice, using the ergodic property for the turbulent phenomena \citep{Tatarsky1967}, it is replaced by averaging over time.

For astronomical applications, the telescope entrance pupil $D$ is considered as a receiving aperture. The scintillation noise can be expressed in magnitudes $\sigma^2_m$. It then becomes the additive noise, associated with the scintillation index: $\sigma^2_m = 1.179\,s^2$.

In nighttime conditions, suitable for astronomical observations, the OT is usually such that the phase and amplitude distortions at the pupil plane are well described under the approximation of weak perturbations \citep{Tatarsky1967}. It is believed that for large apertures (geometric optics regime), this is even more so. In this approximation, the turbulence of each layer is independent of the previous layers, and the scintillation index observed on the surface, is linearly related to the distribution of the structural coefficient $C_n^2(z)$ on the line of sight throughout the whole atmosphere \citep[see, e.g.,][]{Roddier1981} by the following integral:
\begin{equation}
s^2 = \int_A C_n^2(z)\,Q(z) \,{\rm d}z,
\label{eq:s2int}
\end{equation}
where $z$ is the distance to the turbulent layer, which in the case of measurements at zenith, coincides with the altitude $h$ above the observatory. The {\it weighting  function}  $Q(z)$ relates the output from the layer to its corresponding effect at the surface.

\subsection{Basic relations}

The calculation of the weighting function $Q(z)$ has been described in detail in \citep{Tokovinin2002b,Tokovinin2003}. It involves the integration of the 2D spatial spectrum of the amplitude perturbation over all frequencies. In the case of axial symmetry of integrand functions, it is easier to perform the integration in polar coordinates, where, after averaging over polar angle, the functions depend only on the modulus $f$ of the vector of spatial frequencies. For isotropic and locally homogeneous OT with a spatial spectrum of the wavefront phase fluctuations  $\Phi(f)$, normalized to the value
of $C_n^2(z)$:
\begin{equation}
Q(z) = 9.61\int_0^\infty \Phi(f)\,S(z,f) A(f)\,f\,{\rm d}f,
\label{eq:qnorm}
\end{equation}
where the Fresnel filter $S(z,f)$ describes the evolution of the amplitude perturbations in the propagation of the light wave. In the case of monochromatic radiation with wavelength $\lambda$, the filter $S(z,f) = \sin^2(\pi\lambda zf^2)/\lambda^2$, and its intrinsic spatial scale is defined by the Fresnel radius $r_\mathrm{F} = (\lambda z)^{1/2}$.

Aperture filter $A(f)$ takes into account the averaging of wave amplitude by a receiver. For a circular entrance aperture $D$ it is the Airy function $(2J_1(\pi Df)/(\pi Df))^2$ if there are no other factors for averaging. However, in general, stellar scintillation are registered as temporary fluctuations in light intensity averaged over a measurement (exposure). Dependence of the scintillation on the exposure was investigated repeatedly \citep[e.g.,][] {Young1967,Dravins1998} with the help of time averaging the signal.

On the basis of the ``frozen turbulence'' hypothesis \citep{Taylor1938}, the temporal averaging can be replaced with a spatial filtering of the scintillation spectrum \citep[see, e.g.,][]{Martin1987,Tokovinin2002b,2011AstL}, extending the concept of aperture filtering. The scintillation power in the case of a finite exposure is still described by the expression (\ref{eq:qnorm}) if instead of $A(f)$ we substitute the product $A(f)\,A_s(f)$, where $A_s(f)$ is the wind shear filter \citep{Tokovinin2002b,2011AstL}. The real ``freeze'' of the OT is not required as the invariability of the spatial spectrum is enough.

In the cases where measurements are performed with an exposure $\tau$ so short that the wind $w$ in the atmosphere, does not shift the OT by a significant distance $\tau w \ll D$, the additional averaging can be neglected (filter $A_s(f)$ is much wider than the filter $A(f)$ and multiplication does not change the integrand). Hereafter, this situation is called  {\it short (zero) exposure} regime (SE).

In real astronomical observations, the opposite situation is much more common: during the exposure, the wind shifts the OT by distances exceeding the aperture of the telescope $\tau w \gg D$, and temporal averaging becomes the dominant effect (filter $A_s(f)$ is narrower than $A(f)$ and it defines the integrand). This case is denoted below as {\it long exposure} regime (LE).

\subsection{Large apertures approximation}
\label{sec:large_ap}

The expression (\ref{eq:qnorm}) is valid for arbitrary parameters, but in a general form it can be integrated only numerically. To study the dependence of $Q(z)$ on distance $z$ and other parameters, it is necessary to perform some simplification.

To describe the scintillation on a typical telescope, the approximation of {\it large aperture} $D \gg r_\mathrm{F}$ can be used. This approximation is well satisfied for telescopes with diameters of 1~m or more, since for any reasonable atmospheric condition, $r_\mathrm{F} \lesssim 0.1$~m in the optical and near-IR spectral range.

In this case, one can use a well-known feature of the spectral filters included in the integral (\ref{eq:qnorm}): the filters $A(f)$ and $S(f)$ overlap a little. After replacing $\sin(\pi\lambda zf^2)$ on $(\pi\lambda zf^2)$, the integrand is still largely unchanged \citep{Roddier1981}. As a consequence, the dependence on $\lambda$ disappears not only for monochromatic case, but in the case of a wide spectral band of detector, so the more accurate description of this filter from \citep{Tokovinin2003} is not necessary.

Passing to the dimensionless frequency $q = fD$, we can write for circular aperture and the SE regime:
\begin{equation}
Q_\mathrm{S}(z) = 38.44\,D^{-7/3}z^2\int_0^\infty \Phi(q)\,q^3\,(J_1(\pi q))^2{\rm d}q,
\label{eq:qeq_short}
\end{equation}

In the LE regime, the approximation technique is particularly suitable because the wind smoothing additionally suppresses the high-frequencies. We use the asymptotic behavior for the corresponding filter of the wind shear $A_s = D/\pi\tau wq$ from \citep{2011AstL}. Multiplying the integrand by it, we obtain:
\begin{equation}
Q_\mathrm{L}(z) = 12.24\,D^{-4/3}z^2\,(\tau w)^{-1}\int_0^\infty \Phi(q)\,q^2\,(J_1(\pi q))^2{\rm d}q,
\label{eq:qeq_long}
\end{equation}

It should be noted again that the conventional dependencies on telescope diameter $D$ and propagation distance  $z$ are the result of the the Fresnel filter approximation under the condition $D \gg r_\mathrm{F}$ and are not associated with a particular form of the OT spectrum.

For the Kolmogorov model of normalized spatial spectrum of phase perturbations $\Phi(q) = q^{-11/3}$, the integrals $\mathcal I_\mathrm{S}$ and $\mathcal I_\mathrm{L}$ in the right-hand side of these expressions, are easily calculated and are equal to 0.4508 and 0.8699, respectively (see equations \ref{eq:is1} and \ref{eq:il1}). As a result, the previous expressions reduce to the known dependencies $Q_\mathrm{S}(z) = 17.34\,D^{-7/3}z^2$ for SE regime and $Q_\mathrm{L}(z) = 10.66\,D^{-4/3}z^2(\tau w)^{-1}$ for LE regime. After integration over the whole atmosphere, they lead to the following formulae of observed scintillation noise \citep{Young1967,Gilliland1993,Kenyon2006}:
\begin{equation}
s^2_\mathrm{S} = 17.34\,D^{-7/3}\int_A C_n^2(z)\,z^2 {\rm d}z,
\label{eq:short_s2}
\end{equation}
and:
\begin{equation}
s^2_\mathrm{L} = 10.66\,D^{-4/3}\tau^{-1} \int_A \frac{C_n^2(z)\,z^2}{w(z)}{\rm d}z.
\label{eq:long_s3}
\end{equation}
The integrals in these formulae (the {\it atmospheric moments}) are determined by the particular state of the atmosphere and are the figures of merit when monitoring and/or forecast of conditions for photometry \citep{Kenyon2006,2011AstL}. The matter of this paper is to consider the effects which modify the coefficients before the atmospheric moments.

\section{The impact of the outer scale of turbulence}
\label{sec:outer_scale}

The immediate perception of the scintillation as a result of observations with a naked eye and numerous experiments, performed with small telescopes, have created the stable illusion that the scintillation is a sufficiently high frequency process. The problem of contribution of the high- and low-frequencies in the scintillation has been discussed for a long period \citep[see, e.g.,][]{Young1967}. E.g., in one of the first papers devoted to stellar scintillations \citep{Reiger1963}, the author claimed that turbulence outer scale has no effect on the scintillation intensity, in contrast to the inner scale. The fact that the situation changes radically when the measurements are made on large telescopes, usually goes unnoticed.

Nevertheless, we decided to re-evaluate their relative contribution. Assuming the Kolmogorov spectrum, the fraction of the scintillation power in the frequency interval between 0 and some dimensionless frequency $q$ was calculated. The calculation shows that most of the scintillation power ($\sim 70\%$) passes through the first (main) peak of the filter $A(q)$. In the LE regime, the fraction of the power for this peak is even greater  ($\sim 95\%$). The main peak of the aperture filter is located at the $q < 1.22$, which for large telescopes corresponds to meters scale. It is logical to expect that the distinction of the real OT from the usual Kolmogorov model at such scales, may significantly affect the measured power.

\subsection{Non-Kolmogorov models with outer scale}
\label{sec:models}

Skipping the discussion of the physical meaning of the outer scale of turbulence $L_0$ \citep{Tatarsky1967}, which plays an important role in the generation of turbulence per se, we will consider it simply as an additional parameter in the mathematical description of the spectrum of perturbations in the refractive index. The main purpose of this parameter is to limit the infinite spectral density at frequency $f = 0$  inherent in the Kolmogorov spectrum, or to overcome the divergence in the relationships derived from this spectrum.

In astronomical applications, the outer scale of the OT was studied for a long time in connection with long-baseline interferometry \citep{Davis1995,Avila1997JOSA,Maire2006AA}. The development of adaptive optics on large telescopes also requires the correct description of the spectrum of phase perturbations in the low-frequency region \citep{Conan2003,Tokovinin2007,Martinez2010}.

\begin{figure}
\centering
\psfig{figure=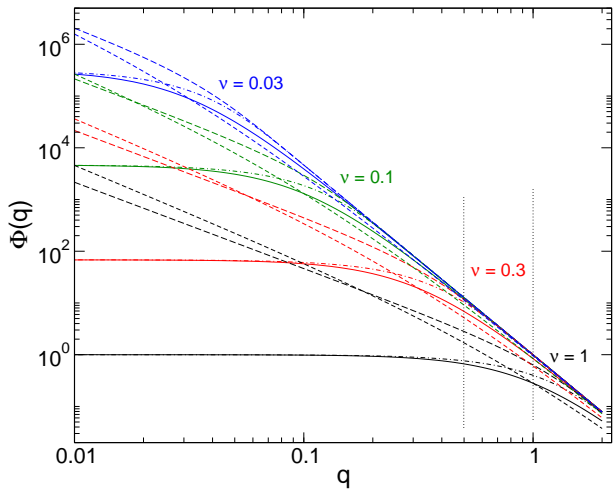,height=8.5cm,angle=-90}
\caption{Spectral power density of the phase perturbations as the function of dimensionless frequency $q$ for different models of non-Kolmogorov turbulence for different values of the dimensionless outer scale $\nu$. Solid lines depict the VK model, short dashed the GT model, long dashed the EM model, dash-dot lines indicate the ME model. See abbreviations in the text. (Online version colored)
\label{fig:spectrums}}
\end{figure}

The most generally accepted OT model which includes the outer scale is the Von Karman (VK) model \citep{vonKarman1948,Tatarsky1967}. In this model, the normalized spectrum of phase perturbation can be written as:
\begin{equation}
\Phi(f,L_0) = (f^2+L_0^{-2})^{-11/6}. % = f^{-11/3}[1+L_0^{-2}f^{-2}]^{-11/6}.
\label{eq:vkarman}
\end{equation}
The main feature of the VK model is a saturation of the power density in the range $f \lesssim L_0^{-1}$ at the level of
$L_0^{11/3}$.

An alternative model without saturation but with a change in exponent of power law, was proposed by Greenwood and Tarazano (GT) \citep{GreenwoodTarazano1974}:
\begin{equation}
\Phi(f,L_0) = (f^2+f\,L_0^{-1})^{-11/6}. % = f^{-11/3}[1+L_0^{-1}f^{-1}]^{-11/6} .
\label{eq:greenwood}
\end{equation}
It is easy to see that at low-frequencies, the spectrum has an asymptote  $\sim L_0^{11/6}f^{-11/6}$. The intermediate range is very wide, so in this model, the divergence from the Kolmogorov power spectrum become apparent early enough.

The exponential model (EM) behaves similarly, it was considered in \citep{Lukin1981AO} as follows:
\begin{equation} \Phi(f,L_0) =
f^{-11/3}[1-\exp(-f^2L_0^2)]. \label{eq:lukin}
\end{equation}
This spectrum  is asymptotic to $\sim L_0^{2}f^{-5/3}$ at low frequencies, ie, has a slightly smaller slope than the GT models. The intermediate range of the spectrum is quite narrow.

We modified the EM  model so, that the spectral density at the origin is finite and coincides with $L_0^{11/3}$, predicted by the VK model. This exponential model (ME) is described by the dependence
\begin{equation}
\Phi(f,L_0) = f^{-11/3}[1-\exp(-f^{11/6}L_0^{11/6})]^{2}.
\label{eq:modif}
\end{equation}
The only difference between this spectrum and VK model is a faster transition from one asymptotic branch to another.

The behavior of the spectral densities for these four models are shown in Fig.~\ref{fig:spectrums} as a function of dimensionless frequency $q = fD$ for different dimensionless outer scale frequency $\nu = D/L_0$. The vertical lines indicates the frequency domain where the main peak of the aperture filter is located. Note that for any model, the spectral density can be described as a product of Kolmogorov spectrum by some spectral filter suppressing low frequencies.

The question of which model best describes the OT in the atmosphere is still open, despite many experimental attempts to solve it \citep[e.g.,][]{Maire2008MN, Wheelon2007ApJS}. For this reason, in further calculations, we use all of the above model with sufficient variety of behavior.

\subsection{The effect of the outer scale on scintillation power}
\label{sec:scint_power_outer}

The impact of the outer scale can be achieved by substituting the appropriate model in the formulae (\ref{eq:qeq_short}) and (\ref{eq:qeq_long}) expressed as a function of dimensionless frequency $q = fD$ and outer scale $\nu = D/L_0$. It is obvious that after such replacement as well as in the case of the Kolmogorov spectrum, dependence on the parameters $z$ and $D$ can be taken out of the integral completely. In the case of the SE regime we obtain:
\begin{equation}
\hat Q_\mathrm{S}(z,\nu) = 38.44\,D^{-7/3}z^2\int_0^\infty \Phi(q,\nu)\,q^3\,(J_1(\pi q))^2{\rm d}q.
\label{eq:qkarm1}
\end{equation}
A similar expression can be written for $\hat Q_\mathrm{L}(z, \nu)$ in the case of LE regime. The value of the integral in the formula depends explicitly only on the parameter $\nu$. However, if the outer scale $L_0$ varies with altitude, then there is an implicit dependence of the integral  on the altitude and the diameter of telescope via the parameter $\nu$.

The approximation for large aperture (\ref{eq:qeq_short}, \ref{eq:qeq_long}) can be used if not only the $D \gg r_\mathrm{F}$ but also $L_0 \gg r_\mathrm{F}$. Otherwise, the integrals diverge and a formula with the exact expression for the Fresnel filter should be used. In that case the integrand maximum is located near $\sim 1/r_\mathrm{F}$ and the situation becomes similar to the scintillation in a small aperture.

Multiplying and dividing the right side of (\ref{eq:qkarm1}) by the value of the integral $\mathcal I_\mathrm{S}$ we get that $\hat Q(z,\nu) = Q(z)\mathcal G(\nu )$, where the function $\mathcal G(\nu)$ describes the impact of the outer scale, and in SE regime is equal to
\begin{equation}
\mathcal G_\mathrm{S}(\nu) = \mathcal I^{-1}_\mathrm{S}\int_0^\infty \Phi(q,\nu)\,q^3\,(J_1(\pi q))^2{\rm d}q.
\label{eq:qdef}
\end{equation}
The smaller this term is, the greater is the effect. The possible factorization of the weighting functions leads to the fact that the scintillation index $\hat s^2$ for non-Kolmogorov OT connects with the index $s^2$, defined by the formulas (\ref{eq:short_s2}) or (\ref{eq:long_s3}) by the simple relationship: $\hat s^2 = s^2 \mathcal G(\nu)$.

In principle, for exponential models the integral (\ref{eq:qdef}) can be calculated analytically, but the resulting expression is so cumbersome that it is meaningless. For VK and GT models we have failed to get an analytic solution, and the usage of expansions of the type \citep{Maire2007MN} leads to the appearance of improper integrals.

\begin{figure}
\centering
\psfig{figure=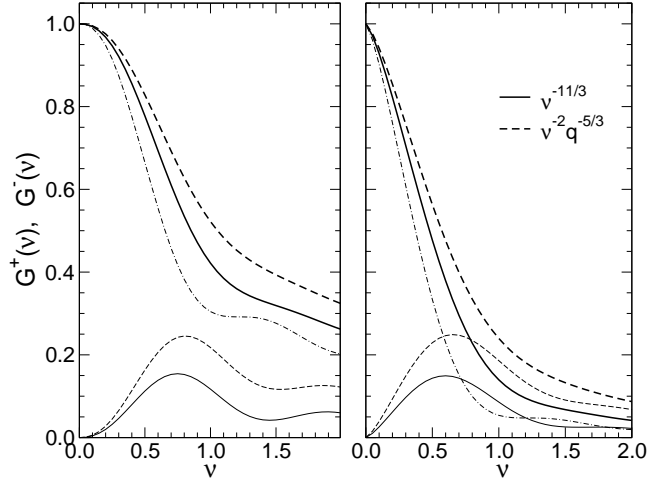,height=8.5cm,angle=-90}
\caption{Dependences of the integrals $\mathcal G^-(\nu)$ (thin solid and dashed lines) and $\mathcal G^+(\nu)$ (dot-dashed line) and the sum of $\mathcal G(\nu)$ (thick lines) on the dimensionless frequency $\nu$ for the piecewise model. Solid lines depict the saturated model, the dased the non-saturated model. On the left: SE regime, on the right: LE regime.
\label{fig:segmen}}
\end{figure}

We therefore investigated an overall behavior of $\mathcal G(\nu)$ using the piecewise power-law dependence with a break point at a frequency $\nu$. The necessary integrals are calculated analytically within two segments of $\{0,\nu\}$ and $\{\nu,\infty\}$. %The particular features of the models will further be analysed on the base numerical integration of the exact equations. * I don't what you want to say in this last sentence *
We will refer to the integral with the spectrum of $q^{-11/3}$ in the range $\{\nu,\infty\}$ as $\mathcal G^+(\nu)$ after its normalization with $\mathcal I$. The integral with the spectrum $\nu^{-11/3}$ (the model with saturation) or $\nu^{-2}q^{-5/3}$ (the model without saturation) in the range $\{0,\nu\}$ will be referred to as $\mathcal G^-(\nu)$. Those constants provide continuity to the spectrum at $q = \nu$. We will specify the SE or LE regimes,  using the appropriate subscript. The integration results of these functions are given in Appendix~\ref{sec:ilimited}.

The functions $\mathcal G^-(\nu)$ and $\mathcal G^+(\nu)$ are shown in Fig.~\ref{fig:segmen} for both regimes. It is evident that the decrease in scintillation power caused by the outer scale, arises from a sharp fall in the function $\mathcal G^+(\nu)$, which is not compensated by an increase in $\mathcal G^-(\nu)$. The figure also shows their sum, i.e. dependence of $\mathcal G(\nu)$. At the point $\nu \approx 0.5 $ the power falls at $\approx 20\%$. As expected, the effect in the case of saturated spectrum is slightly more than for the non-saturated one.

The asymptotes of the $\mathcal G_\mathrm{S}(\nu)$ for small $\nu$ follows from the evaluated integrals in Appendix~\ref{sec:ilimited} and properly describes the functions  while $\nu \lesssim
0.4$:\begin{align}
\mathcal G_\mathrm{S}(\nu) &= 1 - 1.433\,\nu^{7/3} + 1.428\,\nu^{13/3} + \dots, \quad \Phi=\nu^{-11/3} \notag\\
\mathcal G_\mathrm{S}(\nu) &= 1 - 1.083\,\nu^{7/3} + 0.984\,\nu^{13/3} + \dots, \quad \Phi=\nu^{-2}q^{-5/3}
\end{align}
For large $\nu$, these functions come to the asymptotes $0.412\,\nu^{-2/3}$ and $0.508\,\nu^{-2/3}$, respectively.

In the LE regime, the effect of the outer scale is much stronger: at $\nu \approx 0.5$ the power is reduced by almost half and at $\nu \approx 1$ about 5 times. The approximation of the functions $\mathcal G_\mathrm{L}(\nu)$ looks as:
\begin{align}
\mathcal G_\mathrm{L}(\nu) &= 1 - 1.560\,\nu^{4/3} + 1.099\,\nu^{10/3} + \dots, \quad \Phi=\nu^{-11/3} \notag \\
\mathcal G_\mathrm{L}(\nu) &= 1 - 1.276\,\nu^{4/3} + 0.787\,\nu^{10/3} + \dots, \quad \Phi=\nu^{-2}q^{-5/3},
\end{align}
and approximates them into region $\nu \lesssim 0.4$ with an accuracy better than 0.02. In the $\nu \gg 1 $, $\mathcal
G_\mathrm{L}(\nu)$ tend to $0.128\,\nu^{-5/3}$ and $0.375\,\nu^{-5/3}$, respectively.

The difference between the behavior of the functions evaluated for the spectra with and without saturation is not fundamental and can hardly be detected in actual measurements. As we are to see from calculations using the models described in section~\ref{sec:models}, a much greater effect occurs because of the length of intermediate region in these models.

\begin{figure}
\centering
\psfig{figure=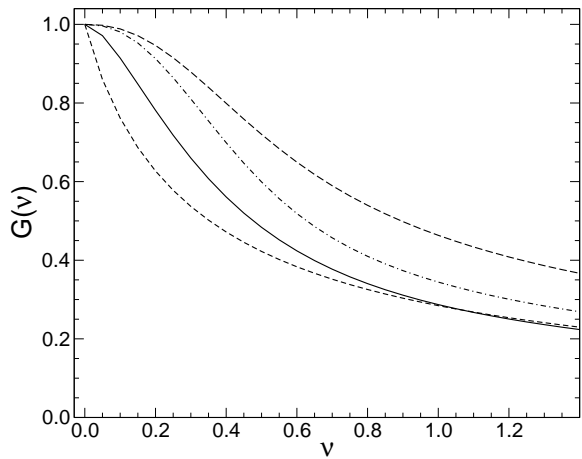,height=8.0cm,angle=-90}
\caption{Dependence of the $\mathcal G(\nu)$ on the dimensionless outer frequency $\nu$ for different models of non-Kolmogorov turbulence for SE regime. Solid line depcts the VK model, short dashed for the GT model, long dashed for the EM model, dash-dot lines indicates the ME model.
\label{fig:short_dep}}
\end{figure}

\begin{figure}
\centering
\psfig{figure=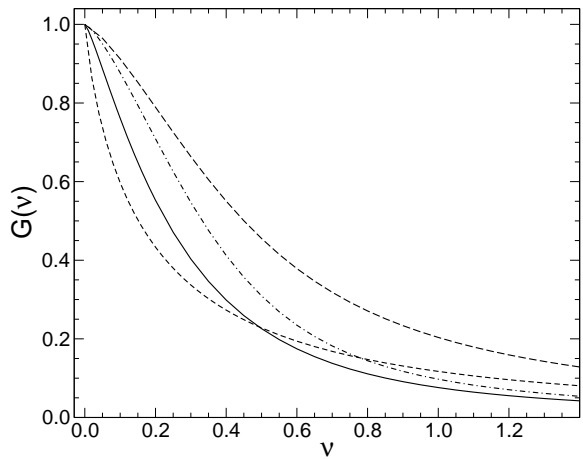,height=8.0cm,angle=-90}
\caption{Dependence of the $\mathcal G(\nu)$ on the dimensionless outer frequency $\nu$ for different models of non-Kolmogorov turbulence for LE regime. See designations in Fig.~\ref{fig:short_dep}.
%Solid line depcts the VK model, short dashed for the GT model, long dashed for the EM model, dash-dot lines indicates the ME model.
\label{fig:long_dep}}
\end{figure}

Fig.~\ref{fig:short_dep} and Fig.~\ref{fig:long_dep} show the functions $\mathcal G(\nu)$, calculated by numerical integration for different non-Kolmogorov models. It is seen that the initial parts ($\nu \lesssim 0.2$) of the curves differ greatly in both SE and LE regimes. E.g., at the point $\nu = 0.1$, the effect is almost zero for the EM model but for the GT model, which has the longest transition region, the function $\mathcal G(\nu) \approx 0.75$. In the LE regine, the $\mathcal G(\nu)$ is 0.92 and 0.60 respectively for these two models. Between these extremes, the curves pass for VK and ME models.

The absolute difference between the models decreases with increasing $\nu$. For models with saturation, the curves pass below the others as seen in Fig.~\ref{fig:long_dep}. A comparison with Fig.~\ref{fig:segmen} shows that the piecewise power-law model approximates a general shape and, having zero transition range, can be considered as a lower estimate of the effect (or upper bound for the $\mathcal G(\nu)$ function) for an appropriate type of models.

\section{The impact of the central obscuration}
\label{sec:co}

Most astronomical telescopes have a central obstruction (CO) caused by the secondary mirror, therefore its impact on power and spectrum of the scintillation was repeatedly considered \citep{Young1967,Dravins1998}. However, these studies have been focused more on qualitative assessment. Calculations of the scintillation spectra served mainly to showcase  the presence of the effect. There are much more studies of the effect of various occulting structures on diffraction image in telescope focal plane \citep{Harvey1995}, which mathematically coincides with the aperture filter necessary for us.

In this section, we once again examine this effect in the approximation of large telescopes, and obtain the expression suitable for accurate description in the case of the Kolmogorov spectrum.

\subsection{Central obscuration for Kolmogorov spectrum}

In the presence of the CO, the aperture filter $A(f)$ is described by the well-known expression \citep[see, e.g.,][]{Harvey1995} depending on additional parameter $\epsilon$, which is a ratio of inner diameter of entrance pupil to its outer diameter. In the formula (\ref{eq:qeq_short}) and follow-up, the aperture filter is separated into factors, and the factor $4/\pi^2$ is taken out of the integrals. Therefore, we write only the part $a(q,\epsilon)$ of the filter, which should  replace the $(J_1(\pi q))^2$ in the integrands.
% \begin{equation}
% a(q,\epsilon) = \left(\frac{J_1(\pi q) - \epsilon\,J_1(\pi \epsilon q) }{1-\epsilon^2}\right)^2\!.
% \label{eq:new_af}
% \end{equation}
\begin{equation}
a(q,\epsilon) = \Upsilon\left(J_1(\pi q) - \epsilon\,J_1(\pi \epsilon q)\right)^2\!,
\label{eq:new_af}
\end{equation}
where the accessory parameter $\Upsilon =(1-\epsilon^2)^{-2}$. While virtually no difference within main spectral peak from the aperture filter of a circular aperture, the filter has a significantly greater transmission within the second and subsequent peaks (e.g., for $\epsilon = 0.3$ about 2.5 times). As a result, the scintillation spectrum at high frequencies rises and the total scintillation power increases as well.

\begin{figure}
\centering
\psfig{figure=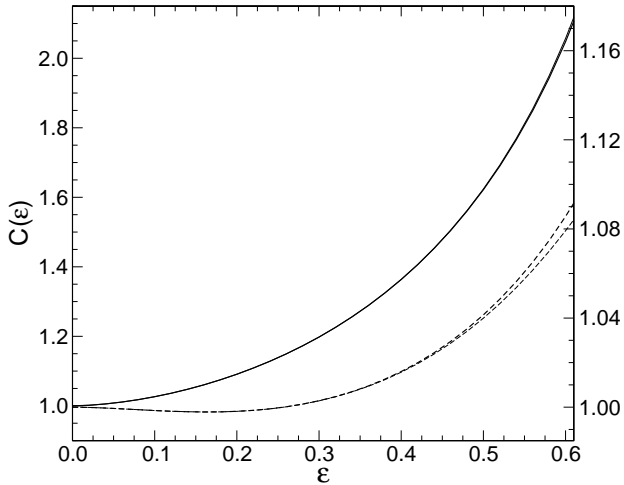,height=8.5cm,angle=-90}
\caption{Relative change in the scintillation power $s^2$ for SE regime (thick solid line, left scale) and for LE regime (dashed line, right scale) on the parameter $\epsilon$. The corresponding thin lines depict the approximations by the formulas \ref{eq:appx_co} and \ref{eq:lco_app}.
\label{fig:co-dep}}
\end{figure}

Substituting the expression (\ref{eq:new_af}) in the (\ref{eq:qeq_short}), we obtain the weighting  function $Q_\mathrm{S}(z,\epsilon)$  which can be represented in the form: $Q_\mathrm{S}(z,\epsilon) = Q_\mathrm{S}(z)\,{\mathcal C}_\mathrm{S}(\epsilon)$, where the integral ${\mathcal C}_\mathrm{S}(\epsilon) $ is expressed in terms of hypergeometric functions as follows:
\begin{equation}
\mathcal C_\mathrm{S}(\epsilon) =  \Upsilon \biggl(1 - \frac{15}{6}\frac{\sqrt[3]{2}\,\Gamma\left(\frac{2}{3}\right)\Gamma\left(\frac{5}{6}\right){}_2F_1\left(\textstyle\frac{1}{6},\frac{7}{6};2;{\epsilon}^{2}\right)}{\sqrt{3\pi} }\,\epsilon^{2} + \epsilon^{5/3}\biggr).
\label{eq:coapprox}
\end{equation}
There is a simple and good approximation for this function. Four members are enough to ensure an accuracy better than 1\% for $\epsilon < 0.6$:
\begin{equation}
{\mathcal C}_\mathrm{S}(\epsilon) \approx \Upsilon (1 + \epsilon^{5/3} - 1.5682\,\epsilon^2 - 0.1525\,\epsilon^4)
\label{eq:appx_co}
\end{equation}
The function ${\mathcal C}_\mathrm{S}(\epsilon)$ is shown in Fig.~\ref{fig:co-dep}. As can be seen, in the SE regime, the CO effect leads to a twofold increase in the power on the telescope with the $\epsilon \approx 0.6$. With widespread CO of $0.3 -  0.4$, the effect is much smaller, though still of $20 - 40\%$.

In the LE regine, the CO effect is calculated similarly using the formula (\ref{eq:qeq_long}). The dependence of the scintillation on the parameter $\epsilon$ is given by the formula
\begin{equation}
\mathcal C_\mathrm{L}(\epsilon) =   \Upsilon\biggl(1-{\frac{8}{27}}\,{\frac{{\pi }
^{3/2}\sqrt [3]{2}\sqrt {3}\,{}_2F_1\left(\textstyle-\frac{1}{3},\frac{2}{3};2;{\epsilon}^{2} \right) }{\Gamma\left(\frac{5}{6}\right)\Gamma\left(\frac{2}{3}\right)}}\epsilon^2 + \epsilon^{8/3}\biggr).
\label{eq:co_long}
\end{equation}
The simple approximation provides accuracy  better than 0.005 in the range $0 < \epsilon < 0.6$ can be written as
\begin{equation}
\mathcal C_\mathrm{L}(\epsilon) \approx \Upsilon (1 - 2.355\,\epsilon^2 + \epsilon^{8/3} + 0.2617\,\epsilon^4)
\label{eq:lco_app}
\end{equation}
The curve in Fig.~\ref{fig:co-dep} shows that in this regime, the CO does not introduce any significant effect for the most of telescopes. Even with the $\epsilon = 0.6$ the power increases only by 8\%. The smallness of the effect is explained by the fact that for the Kolmogorov spectrum, there is some balance between the power decrease in the main peak and an increase in the second peak of the spectrum.

Since the weighting function $Q_\mathrm{S}(z,\epsilon) = Q_\mathrm{S}(z)\,{\mathcal C}_\mathrm{S}(\epsilon)$, all asymptotic dependencies on the altitude and diameter are retained. Accordingly, the effect on the measured scintillation index $s^2$ is also described by the function ${\mathcal C}(\epsilon)$. It is worth noting that if the inner diameter of the aperture $\epsilon D$ becomes comparable with the $r_\mathrm{F}$ (telescope diameter is less than 0.5~m), then the presented above functions are approximate.

\subsection{The combined effect of the outer scale and the central obscuration}

The formulae, obtained in the previous section, are valid if the outer scale effect is negligible, i.e. dimensionless parameter $\nu \ll 1$. Otherwise, the fraction of high-frequency scintillation spectrum grows and the CO effect increases as well.

Obviously, after replacing the aperture filter in the expressions for $\hat Q_\mathrm{S}(z,\nu)$ or $\hat Q_\mathrm{L}(z,\nu)$ (\ref{eq:qkarm1}) on (\ref{eq:new_af}), the integral ${\mathcal  G}(\nu)$ becomes dependent on the parameter $\epsilon$. Separation of it into two multiplicands, each depending on its own parameters, can not be performed. So, the combined impact of these effects can be written either as $\hat Q(z,\epsilon,\nu) = Q(z)\,{\mathcal G}(\nu,\epsilon){\mathcal C}(\epsilon)$ or as $\hat Q(z,\epsilon,\nu) = Q(z)\,{\mathcal G}(\nu){\mathcal C}(\epsilon,\nu)$.

We chose the latter form, because for extremely large telescopes the effect of the outer scale is dominant. Accordingly the function ${\mathcal C}(\epsilon,\nu)$ depends on the adopted model and not only on the parameter $\nu$. The normalization, used in these calculations, provides that ${\mathcal C}(\epsilon,\nu) = {\mathcal C}(\epsilon)$ when $\nu = 0$.

\begin{figure}
\centering
\psfig{figure=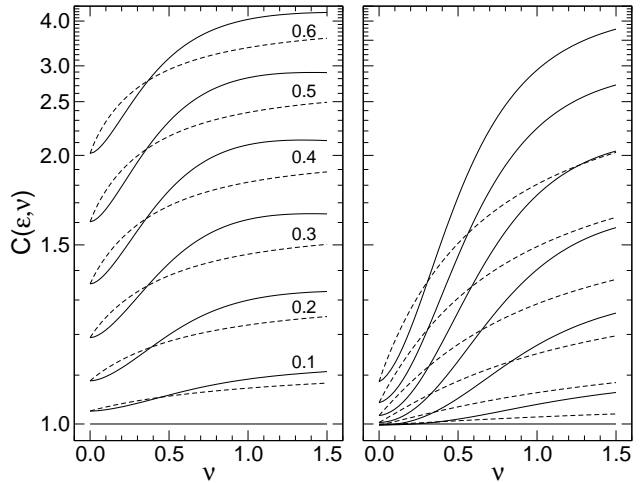,height=8.5cm,angle=-90}
\caption{Functions ${\mathcal C}(\epsilon,\nu)$ for $\epsilon = 0, 0.1, 0.2, 0.3, 0.4, 0.5, 0.6$ depending on $\nu$ for SE (on left) and LE (on right) regimes. The solid lines  depict VK model, dashed lines the GT model.
\label{fig:co-common}}
\end{figure}

The curves ${\mathcal C}(\epsilon,\nu)$, calculated for the VK and GT models, are shown in Fig.~\ref{fig:co-common}. For both models the CO effect increases monotonically with $\nu$. It is seen, that for the VK model, which has a saturation at low spatial frequencies, the effect is more significant and at $\nu \approx 1$ almost reaches its maximum. This indicates that the scintillation within the main peak of the aperture filter, is suppressed almost completely.

In the LE regime, the behavior of ${\mathcal C}(\epsilon,\nu)$ is also very interesting. For small $\nu$ the CO effect is negligible on real telescopes, but at $\nu \sim 0.5$ becomes comparable with the effect in the SE regime and is growing further. Nevertheless this does not compensate for the scintillation reduction caused by the outer scale effect (Fig.~\ref{fig:long_dep}). The VK and GT models differ more than in the SE regime.

\section{Discussion}
\label{sec:discuss}

\subsection{Scintillation noise on large and and extremely large telescopes}

\begin{figure}
\centering
\psfig{figure=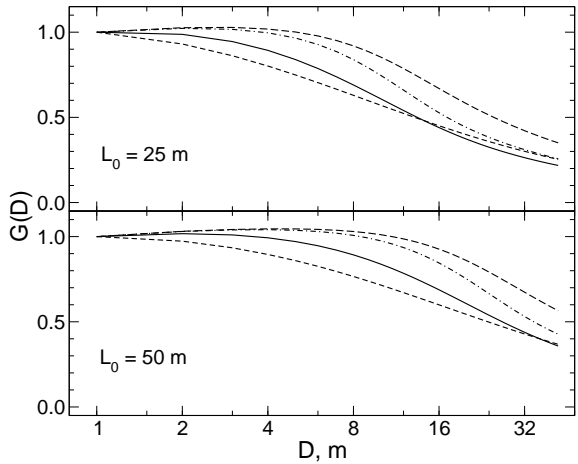,height=8.0cm,angle=-90}
\caption{ Effect of the outer scale as a function of the diameter of telescope for two values of $L_0$ for SE regime. The curve for the VK model is shown with the solid line, for GT model with the dashed, for EM with long dashed and for ME model with dash-dotted line.
\label{fig:short-D}}
\end{figure}

The value of the outer scale is important for many applications, so it is often measured by different methods and its estimates and altitude profiles are available for several astronomical observatories \citep{Abahamid2004,Maire2007MN,DaliAli2010,Floyd2010}. These data show that at altitudes greater than 8~km, the typical $L_0$ is $20 - 25$~m, although values from 10 to 100~m are sometimes observed. We will use these figures for further numerical evaluation.

\begin{figure}
\centering
\psfig{figure=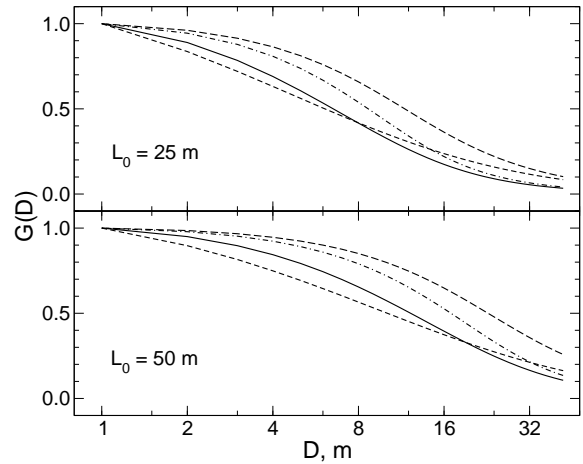,height=8.0cm,angle=-90}
\caption{Effect of the outer scale as a function of the diameter of telescope for two values of $L_0$ for LE regime.
 See designations in Fig.~\ref{fig:short-D}.
%The curve for the VK model is shown with the solid line, for GT model with the dashed, for EM with long dashed and for ME model with dash-dotted line.
\label{fig:long-D}}
\end{figure}

Fig.~\ref{fig:short-D} and Fig.~\ref{fig:long-D} show $\mathcal G(\nu)$ as a function of the telescope diameter for $L_0=25$~m and $L_0=50$~m. Here the CO effect is not considered, so the curves are applicable only to telescope with $\epsilon \lesssim 0.2$. We see that even in SE regime on telescopes of 10~m class, the scintillation power is $\approx 0.6$ from the power predicted by the standard relation (\ref{eq:short_s2}), assuming the actual spectrum of perturbations close to the VK or GT models. Exponential models lead to a smaller effect.

For the designed extremely large telescopes TMT and E-ELT, the $\mathcal G$ decreases to $\approx 0.25$, that certainly is to be taken into account in the estimates of errors budget in high-precision photometric observations. Note that the CO for the TMT ($\epsilon = 0.11$) and for the E-ELT ($\epsilon = 0.15$) is small, so that its accounting increases corrective functions by no more than 10 \%.

For the LE regime, the effect is more significant: the measured scintillations power is reduced by  $10 - 20$ times. Using the formula (\ref{eq:long_s3}) and typical estimates of atmospheric moments from \citep{Kenyon2006}, we can predict that for 60~s exposure the scintillations noise becomes $\approx 10~\mu\mbox{mag}$ instead of $\approx 40~\mu\mbox{mag}$ predicted by the classical formula. For comparison, the similar estimate for 4~m telescope accounts  $\approx 200~\mu\mbox{mag}$.

Naturally, the scintillation noise at the level of $10~\mu\mbox{mag}$ requires an appropriate photon noise. It is easy to calculate that $10^{10}$ photons are to be accumulated at least, or, in the case of TMT, the source should provide illumination about $25\mbox{ photons\,cm}^{-2}\mbox{s}^{-1}$, which corresponds to a star with $V \approx 11.5$ magnitude. Of course, for such photon fluxes, special technique should be used to avoid strong nonlinearity effects.

Evaluation of the scintillation impact in the case of fast photometry (on time scales $0.001-0.01$~s) for the same telescope, leads to the noise as $100-150~\mu\mbox{mag}$, which is about 20 times less than for 4~m telescopes. The brightness of objects available for such measurements, depends on the exposure and for $\tau = 0.01$~s is equal to $\approx 7.0$ magnitude.

\begin{figure}
\centering
\psfig{figure=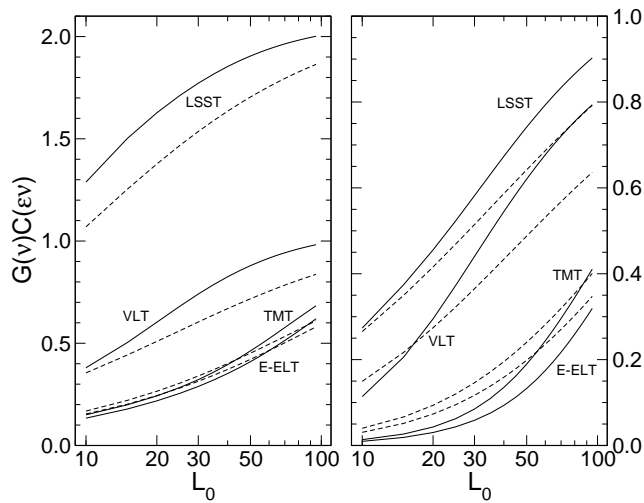,height=8.5cm,angle=-90}
\caption{The combined effect of the outer scale and the CO as a function of $L_0$ for 3 designed telescopes: LSST, TMT, E-ELT, and for the telescope VLT. The curves for the VK model are shown with solid lines, for GT model with  dashed lines. The case of SE regime is shown on the left, the LE regime on the right.
\label{fig:teles}}
\end{figure}

The combined effect of the outer scale and the CO for two large (VLT and LSST) and two extremely large (TMT and E-ELT) telescopes is illustrated in Fig.~\ref{fig:teles}, where the effect is represented as a function of the outer scale $L_0$.  Evidently, the situation for the telescope LSST ($D = 8.36$ ~m) with a large $\epsilon = 0.61$  is distinguished.  In the SE regime, the scintillation is larger than estimate gives by the formula (\ref{eq:short_s2}), since CO effect dominates over the outer scale effect. However, in the LE regime and for the most probable value of $L_0 = 25$~m the scintillation power is expected half as many.

The complex rarefied configuration of aperture of the GMT telescope has no evident diameter, so we can not obtain an estimate of the scintillation power in the usual way. Instead, we use the fact that the scintillation is almost uncorrelated at distances greater than the diameter of a telescope \citep{2012aMNRAS}. In the LE regime, the correlation can be as small as ($\sim 0.2$), but owing to outer scale effect it should be further reduced.

Ignoring the CO, although it for the central GMT mirror is large enough, we find that the scintillation power on  GMT should be by $4-7$ times smaller than for a single mirror with diameter 8.36~m. Therefore the curves for the telescope VLT, shown in Fig.~\ref{fig:teles}, can be used as an estimation.

\subsection{General questions}

All the extremely large telescopes are designed with segmented primary mirror. The gaps between the individual segments and secondary mirror spiders have a characteristic size of about $0.2 - 2$~cm, and also increase the power of high-frequency scintillation. Maximum effect falls within  second -- third transmission peak of the aperture filter as in the case of the CO. In the first approximation, the effect is proportional to obscured relative area. The area of gaps between segments amounts to less than 1\% (for TMT it is 0.3\%), so that even in the SE regime, we does not expect that their influence is noticeable.

%*I would drop this upcoming paragraph. The content is rather obvious  and having this here feels a bit random* Recall that we used the distance to turbulent layer $z$, and not the altitude $h$ above observatory. Since both the effects do not change dependence on $z$, then the behavior of the scintillation power with zenith distance $Z$ (or air mass $M_\mathrm{Z} = \sec Z$) does not differ from the usual: in the SE regime it is $\propto M_\mathrm{Z}^{3}$, and in the LE regime the exponent varies from 3 to 4 depending on the wind direction \citep{Young1967,2011AstL}.

In assessing the observed effect of the outer scale in the section ~\ref{sec:scint_power_outer}, we have assumed that the $L_0$ is constant on the line of sight. Otherwise, the function $\mathcal G(\nu)$ can not be factored out of integrals (\ref{eq:short_s2}) or (\ref{eq:long_s3}) and this effect in the scintillation index should be evaluated for each layer and then integrate over the whole atmosphere. Evidently, it is necessary to know the vertical distribution of OT and $L_0$ for this.

However, the calculated effect is model-dependent and the imprecision, introduced by the uncertainty of the model, is quite large (see Fig.~\ref{fig:short_dep} and Fig.~\ref{fig:long_dep}), so variation of $L_0$ with altitude on the order of $\pm 50\%$ can be neglected. The second circumstance, which has already been mentioned: in large, the effect is defined by the $L_0$ in a bounded region of the altitudes ($10 - 15$~km) which generates the main part of the scintillations for large telescopes.

The curves in Fig.~\ref{fig:teles} also explain that the temporal variations of the measured scintillation power are to occur not only because of changes in OT intensity in the upper atmosphere, but also because of variations in the outer scale. The net effect should be greatly depended on how these two parameters are connected. This question is virtually unexplored, but it is clear that if an increase in the intensity correspond to a decrease of the $L_0$, then the scintillation power is to be quite stable characteristic, and vice versa.

The functions, shown in Fig.~\ref{fig:short-D} and~\ref{fig:long-D}, can be used for verification of type of the turbulence models. For this, the initial part of the dependencies (diameters of $2 - 8$~m) for the SE regime is best suited. Here, simultaneous measurements with apertures of different diameters must be required. One can apply the known methods of integration of exit pupil image, selecting the desired sub-apertures during processing. In order to minimize an impact of the CO, this sub-apertures should be similar. E.g., on the VLT telescope one can distinguish a number of sub-apertures with a diameter of 8.2~m, 3.5~m and less with the same $\epsilon = 0.15$.

Reduction of low-frequency components of the scintillation for non-Kolmogorov OT leads to yet another consequence: an effectiveness of the pupil apodization, proposed in \citep{Young1967}, should greatly increase. In the case of extremely large telescopes, it is expected that optimal apodization should lead to a gain more than 2 times in the scintillation noise.

% 1. Turbulence description by a stationary random process probably does not
% work well at large scales. The calculations are still useful as a guideline
% (dependence on D, z), but should be regarded as estimates (the apparent
% accuracy is misleading).
%
% 2. Same for the Taylor hypothesis. Turbulence evolution will probably add
% some averaging, so the use of Taylor may give upper limit on scintillation.
% It's worth to comment on both aspects.
%
% 3. Standard theory works on large apertures well even for saturated
% scintillation (it is essentially geometric optics). Substantial reduction of
% scintillation with outer scale is a problem in lunar/solar scintillometer. At
% z=10km, the cone is 100m across. We take L0 in consideration in the
% calculation of LuSci weighting functions.
%
% 4. If central obstruction is significant, so may be other "details" of the
% aperture function such as spiders and gaps in the segmented mirror. Will those
% dominate in the short-exposure case? These details will also change the image
% structure (diffraction rays etc.), so the problem of estimating photometric
% error  in PSF-fitting photometry becomes quite complicated.
%
% 5. GMT with large gaps between 8-m mirrors is a special case. Worth mentioning?
% It should be in-between filled 22m and 8m somewhat attenuated by averaging 7
% beams.
%
% 6. The 10umag estimate for TMT opens an exciting science niche: look for
% transits of earth-like planets in clusters.

\section{Conclusion}
\label{sec:conclusion}

In this paper we considered two factors that modify the scintillation intensity in observations on large ($4 - 10$~m) and extremely large ($20-50$~m) telescopes. The first effect is caused by deviation of the real OT from the Kolmogorov model at the scales of the order of 10~m. In the models which describe the real turbulence with using outer scale $L_0$, the scintillation power at low spatial frequencies is much lower than for pure Kolmogorov spectrum.

Effect of the outer scale can be described by an additional function, which depends on the ratio of $D/L_0$. This function is equal to 1 when the telescope aperture is small and decreases with diameter increasing. We considered its general behavior by the example of piecewise power-law spectrum with the salient point at spatial frequency $L_0^{-1}$. Particular features were investigated by numerical integration for four different models: Von Karman and Greenwood-Tarazano models, and two exponential models.

For the observed values of $L_0$ and models with a wide intermediate zone, the effect becomes visible on 4~m class telescopes. For measurements with long exposures this effect is more important than in the case of short exposures, and for the extremely large telescopes TMT and E-ELT it can reduce the scintillation power $\sim 10$ times compared to classical estimates.

For the Kolmogorov OT, the effect of amplification of the scintillation power, caused by the CO, inherent in every large telescope, results in the multiplication of the power for circular aperture by the function that depends only on the CO parameter. In the case of models with outer scale, the effect is described by a more complicated way, and it becomes significant for large apertures for both short and long exposures.

The significant reduction of scintillation noise, due to the outer scale, enhances the potential of ground-based telescopes to study the variability of astronomical objects at $\sim 10^{-5}$ level. This accuracy is sufficient to see, e.g., transit of Earth-like planet across the disk of Solar-like star. Drastic increase in the accuracy of fast photometry (up to $\sim 10^{-4}$) makes it possible to study the micro-variability of many astronomical objects without the accumulation of long time series suitable for temporal spectral analysis.

\section{Acknowledgements}

The author thanks his colleagues and in particular B.~Safonov and A.~Tokovinin for valuable comments and suggestions during the discussion of this work. He is especially grateful to T.~Travouillon, whose corrections significantly improved the paper.

\bibliography{outer_eng}
\bibliographystyle{mn2e}

\appendix
\section{Integrals}
\subsection{Integrals in the case of Kolmogorov spectrum}
Here the known integrals, needed for calculating the weighting function, are listed with the formula (\ref{eq:qeq_short}) in SE regime:
\begin{multline}
\mathcal I_\mathrm{S} = \int_0^\infty q^{-11/3}\,q^3\,(J_1(\pi q))^2{\rm d}q \\=  \frac{2\,\pi^{7/6}\sqrt{3}}{15\,\Gamma\!\left(\textstyle\frac{2}{3}\right)\Gamma^3\!\left(\textstyle\frac{5}{6}\right)} = 0.4508,
\label{eq:is1}
\end{multline}
and with formula (\ref{eq:qeq_long}) in LE regime:
\begin{multline}
\mathcal I_\mathrm{L} = \int_0^\infty q^{-11/3}\,q^2\,(J_1(\pi q))^2{\rm d}q \\= \frac{27\,2^{2/3}\sqrt{3}\,\Gamma\!\left(\textstyle\frac{2}{3}\right)\Gamma^2\!\left(\textstyle\frac{5}{6}\right)}{32\,\pi^{4/3}} = 0.8699.
\label{eq:il1}
\end{multline}

\subsection{Integrals in the case of finite limits}
\label{sec:ilimited}
In this appendix, the results of the integration over two spectral regions delimited by the dimensionless frequency $\nu$ are written for the cases with a different exponent of the power spectrum of the phase perturbations $\Phi(q)$ (see \ref{sec:scint_power_outer}). Expressions for the SE and LE regimes are marked with the corresponding subscript. The high-frequency part $q > \nu$ with spectrum $\Phi(q) = q^{-11/3}$ looks as
\begin{multline}
\mathcal I_\mathrm{S}^+(\nu) = \int_\nu^\infty q^{-11/3}\,q^3\,(J_1(\pi q))^2{\rm d}q \\= \mathcal I_\mathrm{S} - \frac{3\pi^2}{28}\,\nu^{7/3}\,{}_2F_3\left(\textstyle\frac{7}{6},\frac{3}{2};2,\frac{13}{6},3; -\pi^2\nu^2\right)
\label{eq:short_tail},
\notag
\end{multline}
\begin{multline}
\mathcal  I_\mathrm{L}^+(\nu) = \int_\nu^\infty q^{-11/3}\,q^2\,(J_1(\pi q))^2{\rm d}q = \frac{81\,\Gamma\!\left(\textstyle\frac{5}{6}\right)\Gamma^3\!\left(\textstyle\frac{2}{3}\right)}{32\pi^{11/6}} \\- \frac{3\pi^2}{16}\nu^{4/3}{}_2F_3\left(\textstyle\frac{2}{3},\frac{3}{2};\frac{5}{3},2,3;-\pi^2\nu^2\right).
\end{multline}

Low-frequency part $q < \nu$ in the case of the spectrum with the saturation $\Phi(q) = \nu^{-11/3}$:
\begin{multline}
\mathcal  I_\mathrm{S}^-(\nu) = \int_0^\nu \nu^{-11/3}\,q^3\,(J_1(\pi q))^2{\rm d}q \\=\frac{\nu^{1/3}}{6}{\bigl(\left(J_2(\pi\,\nu)\right)^2 + \left(J_1(\pi\,\nu)\right)^2\bigr)}, \notag
\end{multline}
\begin{multline}
\mathcal  I_\mathrm{L}^-(\nu) = \int_0^\nu \nu^{-11/3}\,q^2\,(J_1(\pi q))^2{\rm d}q \\= \frac{\pi^2}{20} \nu^{4/3}{}_2F_3\left(\textstyle\frac{3}{2},\frac{5}{2};2,3,\frac{7}{2};-\pi^2\nu^2\right).
\label{eq:short_head1}
\end{multline}
Low-frequency part $q < \nu$ in the case of the spectrum without the saturation $\Phi(q)=\nu^{-2} q^{-5/3}$:
\begin{multline}
\mathcal  I_\mathrm{S}^-(\nu) = \int_0^\nu \nu^{-2}q^{-5/3}\,q^3\,(J_1(\pi q))^2{\rm d}q \\= \frac{3\pi^2}{52}\,\nu^{7/3}{}_2F_3\left(\textstyle\frac{3}{2},\frac{13}{6};2,3,\frac{19}{6},-\pi^2\nu^2\right),
\notag
\end{multline}
\begin{multline}
\mathcal  I_\mathrm{L}^-(\nu) =  \int_0^\nu \nu^{-2}q^{-5/3}\,q^2\,(J_1(\pi q))^2{\rm d}q \\= \frac{3\pi^2}{40}\,\nu^{4/3}{}_2F_3\left(\textstyle\frac{3}{2},\frac{5}{3};2,3,\frac{8}{3};-\pi^2\nu^2\right).
\label{eq:short_head2}
\end{multline}
\label{lastpage}

\end{document}